\documentclass[useAMS,usenatbib,onecolumn,usegraphicx]{mn2e}

\newcommand{\modonepwf}{1}
\newcommand{\modtwopwf}{2}
\newcommand{\modthreepwf}{3}
\newcommand{\modfourpwf}{4}
\newcommand{\modfivepwf}{5}
\newcommand{\modsixpwf}{6}
\newcommand{\modsevenpwf}{7}
\newcommand{\modninepwf}{8}
\newcommand{\modtenpwf}{9}
\newcommand{\modonedpn}{10}
\newcommand{\modtendpn}{11}

\title[Neutrino Scattering, Absorption and Annihilation above the accretion disks of Gamma Ray Bursts]{Neutrino Scattering, Absorption and Annihilation above the accretion disks of Gamma Ray Bursts}

\author[J.P. Kneller, G. C. McLaughlin and R. Surman]{J.P. Kneller$^{1}$, G. C. McLaughlin$^{1}$ and R. Surman$^{2}$\\
$^{1}$Department of Physics, North Carolina State University, Raleigh, NC  27695\\
$^{2}$Department of Physics, Union College, Schenectady, NY 12308 }

\begin{document}

\date{Accepted. Received.}

\pagerange{\pageref{firstpage}--\pageref{lastpage}} \pubyear{200x}

\maketitle

\label{firstpage}

\begin{abstract} 
The central engine that drives gamma ray burst (GRB) explosions may derive 
from the ability of electrons/positrons and nucleons to tap into the 
momentum and energy from the large neutrino luminosity emitted by an accretion disk surrounding 
a black hole. This transfer of momentum and energy occurs due to neutrino absorption, 
scattering, and annihilation and the non-spherical geometry of the source 
both increases the annihilation efficiency and, close to the black hole, 
directs the momentum transfer towards the disk axis. 
We present annihilation efficiencies and the 
momentum/energy transfers for a number of accretion disk models 
and compute the critical densities of infalling material below which the transfer of 
neutrino momentum/energy will lead to an explosion. Models in which the neutrinos 
and antineutrinos become trapped within the disk have noticeably different 
momentum and energy deposition structure compared to thin disk models 
that may lead to significant differences in the explosion dynamics. 

\end{abstract}



\section{Introduction}

Gamma ray bursts were first observed forty years ago, but only recently has 
significant progress been made in understanding their origin. Evidence is 
mounting that at least the `long duration' bursts are associated with a 
rare type of supernova event such as a `failed supernova' or `collapsar'
\citep{woo93,pac98,mac99,mac01,prog03}, with neutron-star mergers as a 
candidate for the shorter bursts \citep{pac91,ruf99,ros02}. In either case, it is likely that an accretion disk surrounding a black hole forms that cannot cool efficiently by photon emission because of the high densities and temperatures. But with their much smaller cross sections the neutrinos, if any are produced, may escape and cool the disk. 

The electron neutrinos and antineutrinos emitted from the disk are produced from electron capture by protons or positron capture by neutrons respectively. Both reactions modify the make-up of the disk with electron capture initially predominating as material moves inward toward the black hole. The frequency of these reactions (and their inverse) are strong functions of the thermal properties of the disk and grow considerably as material moves towards the black hole where the temperatures and densities are highest. In turn, the density and temperature are functions of the mass, $M_{BH}$, and spin, $a$, of the black hole, the viscosity, $\alpha$, of the disk and the accretion rate, $\dot{m}$ since these four parameters determine the potential and the rate at which material flows radially inward. In particular, as either $\dot{m}$ or $a$ increase so does the density of the disk (particularly in the hottest region close to the black hole) leading to greater neutrino luminosity. 

The neutrino scattering length within the disk is similarly affected by the thermal properties. If sufficiently high temperatures are reached the neutrinos and/or antineutrinos become trapped and, consequently, their spectrum is altered. Due to the change in nucleon composition as material flows inward, neutrinos and antineutrinos will have different optical depths and will not become trapped at the same radius. For the lower accretion rate disks, $\dot{m} \la 0.1 \,\rmn{M_{\sun} \,s^{-1}}$, the density of the disk is never sufficient to trap the neutrinos and so they immediately escape. But for the higher accretion rates the neutrinos \emph{do} become trapped. Once trapped the neutrinos thermalize and the neutrinos emitted from these regions have temperatures of a few MeV, the exact value depends on the model, and are functions of the radius. In addition, neutrino trapping also has the effect of restricting neutrino emission to that portion of the disk beyond a `neutrinosurface' which, given the cylindrical symmetry, are roughly toroid in shape.   

Descriptions of disks with low accretion rates are given in \citet{pwf99} while models which do include neutrino trapping are given in \citet{dim02}. Additional calculations of neutrino emission from accretion disks are given in \citet{set04}.

After escaping the disk the neutrinos and antineutrinos do not propagate unimpeded but may interact with material above the disk transferring energy and momentum. The extent to which their energy and momentum is tapped is the primary interest of this paper since the imparted energy may\footnote{Though some of the required energy may come by electromagnetic extraction of energy from the rotating black hole, e.g. through the mechanism of \citet{Bla77}}  power the burst. The transfer of momentum and energy occurs by a number of processes that we shall discuss in \S \ref{sec:nuinn}. We take calculations of the neutrino flux emitted from every point on the disk from previous work in \citet{sur04}. Our results are presented in \S \ref{sec:results} and to investigate the effects of each of the four parameters needed to describe the disk we compute the energy/momentum transferred in a number of models. The parameters for each are listed in Table \ref{table:models}. 


\section{Neutrino Interactions} \label{sec:nuinn}

The neutrinos and antineutrinos that emerge from the accretion disk can 
interact in a number of different ways with the material above the disk 
as they propagate. These interactions include
\begin{equation}
\nu_e + \bar{\nu}_e  \rightarrow  e^+ + e^-, \label{reaction:nunubar}
\end{equation}
\begin{eqnarray}
\nu_e + n         & \rightarrow & e^- + p, \label{reaction:nunabs} \\ 
\bar{\nu}_e + p   & \rightarrow & e^+ + n, \label{reaction:nubarpabs}
\end{eqnarray}
\begin{eqnarray} 
\nu_e +e \rightarrow \nu_e' +e' & & \bar{\nu}_e + e \rightarrow \bar{\nu}_e'+e', \label{reaction:nue} \\
\nu_e +p \rightarrow \nu_e' +p' & & \bar{\nu}_e +p \rightarrow \bar{\nu}_e'+p',  \label{reaction:nup} \\
\nu_e +n \rightarrow \nu_e' +n' & & \bar{\nu}_e +n \rightarrow \bar{\nu}_e'+n'.  \label{reaction:nun}
\end{eqnarray}
We discuss each in turn.


\subsection{Neutrino Annihilation}

\begin{figure}
\includegraphics[width=7in]{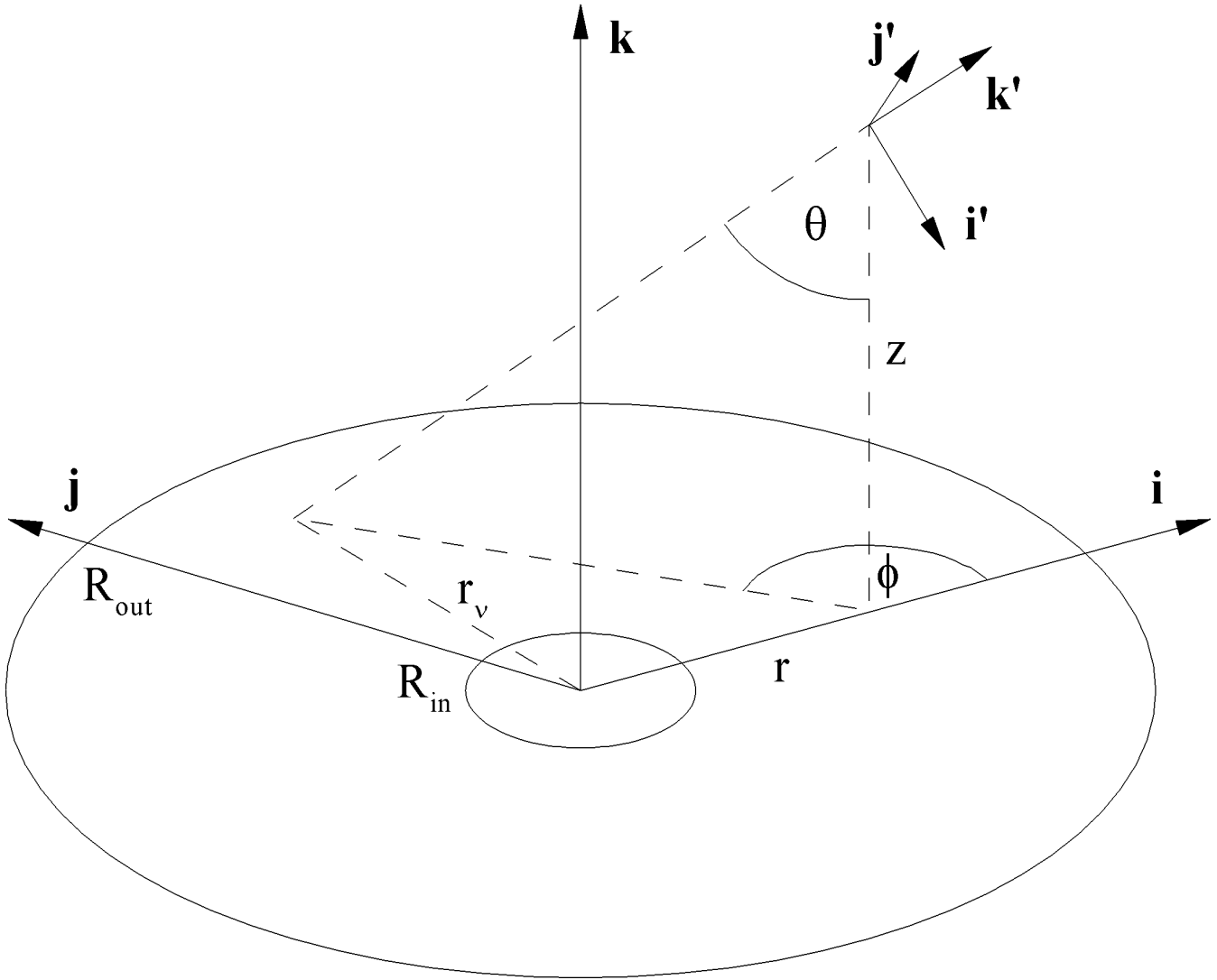}
\caption{Basis vectors, angles and distances used in the calculation of neutrino annihilation, absorption and scattering. The black hole is located at the origin and the accretion disk has inner and outer radii of $R_{in}$ and $R_{out}$. Neutrinos, emitted from the disk at $r_{\nu}$, travel along the vector ${\bf k}'$ toward the point $r,z$ located in the plane formed by ${\bf i}$ and ${\bf k}$. The vector ${\bf i}'$ lies within the plane containing $\theta$.}
\label{fig:geometry}
\end{figure}

Annihilation of neutrino-antineutrino pairs to electron-positrons, reaction (\ref{reaction:nunubar}), will deposit energy at a rate per unit volume, $dL_{\nu\bar{\nu}}/dV$, that is a function of the (cylindrical) coordinates $r$ and $z$, symmetry ensuring there is no polar angle dependence. The expression for $dL_{\nu\bar{\nu}}/dV$ is
\begin{eqnarray}
\frac{dL_{\nu\bar{\nu}}}{dV} & = & \int\int d\Omega_{\nu} \, d\Omega_{\bar{\nu}} \int\int dE_{\nu} \, dE_{\bar{\nu}} \; \frac{f_{\nu}(r_{\nu},E_{\nu})\,f_{\bar{\nu}} (r_{\bar{\nu}},E_{\bar{\nu}})}{16\pi^{2}} \; S_{\nu}(\theta_{\nu},\phi_{\nu})\,S_{\bar{\nu}}(\theta_{\bar{\nu}},\phi_{\bar{\nu}}) \nonumber \\
& & \times \frac{(E_{\nu}+E_{\bar{\nu}})}{E_{\nu}\,E_{\bar{\nu}}} \; \left\{ E_{\nu}\,E_{\bar{\nu}}\; 
|{\bf v}_\nu - {\bf v}_{\bar{\nu}}|\; \sigma_{\nu\bar{\nu}} \right\}. \label{eq:ann_rate}
\end{eqnarray}
The quantities $f_{\nu_e}(r_{\nu},E_{\nu})$ and $f_{\bar{\nu}_e} (r_{\bar{\nu}},E_{\bar{\nu}}) $ are the differential neutrino and antineutrino number fluxes per unit area that emerge from radial coordinates $r_{\nu},r_{\bar{\nu}}$ on the disk. We take these fluxes from \citet{sur04} and we refer the reader to that paper for details of their calculation. 
The two radial positions may be expressed in terms of the angles $(\theta_{\nu},\phi_{\nu})$ and $(\theta_{\bar{\nu}},\phi_{\bar{\nu}})$ together with $r$ and $z$, as shown in Fig. \ref{fig:geometry}, so that 
\begin{eqnarray}
r_{\nu}^{2} & = & \left( r\cos \phi_{\nu} + z\tan \theta_{\nu} \right)^{2} + r^{2}\sin^{2} \phi_{\nu} \\
r_{\bar{\nu}}^{2} & = & \left( r\cos \phi_{\bar{\nu}} + z\tan \theta_{\bar{\nu}} \right)^{2} + r^{2}\sin^{2} \phi_{\bar{\nu}}. 
\end{eqnarray}
The term in curly braces in equation (\ref{eq:ann_rate}) is a Lorentz invariant quantity so can be evaluated in the center of mass frame and expressed in the terms of the invariant Mandelstam variable $s$. When in the CM frame of the collision $s = 4E_\nu E_{\bar{\nu}}$; for the frame of the disk $s = 2 E_\nu E_{\bar{\nu}} ( 1 - \cos \gamma)$ where $\gamma$ is the angle between the $\nu$ and $\bar{\nu}$ trajectories, given by  
\begin{equation}
\cos\gamma = \sin\theta_{\nu}\,\sin\theta_{\bar{\nu}},\cos(\phi_{\nu}-\phi_{\bar{\nu}}) + \cos\theta_{\nu}\,\cos\theta_{\bar{\nu}}.
\end{equation}
The annihilation cross section is well known \citep[see][for example]{HH1989} and so
\begin{equation}
E_\nu E_{\bar{\nu}}\; |{\bf v}_\nu - {\bf v}_{\bar{\nu}}|\; \sigma_{\nu\bar{\nu}} = \frac{\sigma_{0}\,s^{2}}{48\, m_e^2}\;
\sqrt{1-\frac{4 m_{e}^{2}}{s}} \;\left[ 2(C_{V}^{2}+C_{A}^{2}) + \frac{4 m_{e}^{2}}{s}\left(C_{V}^{2} - 2\,C_{A}^{2}\right) \right]
\end{equation}
with $C_{V}=1/2 + 2 \sin^{2}\theta_{W}$, $C_{A}=1/2$ and $\theta_{W}$ being the Weinberg angle, while $\sigma_{0} = 4\,G_{F}^{2}\,m_{e}^{2}/(\pi\,\hbar^{4})$. Finally, the two $S(\theta,\phi)$ terms in equation (\ref{eq:ann_rate}) are corrections to the neutrino and antineutrino differential number fluxes per unit area that are required in order to take into account the fact that neutrinos and/or antineutrinos within the disk may be trapped. When trapping occurs we may divide the disk into optically thick and optically thin zones, the optically thick region being closer to the black hole, with radial boundaries of $a_{\nu}$ and $a_{\bar{\nu}}$ for neutrinos and antineutrinos respectively. The values of $a_{\nu}$ and $a_{\bar{\nu}}$, where appropriate, are listed in table (\ref{table:models}). In doing so we make use of the approximation that the scattering length in the optically thick zone is much smaller than the disk scale height so that the transition from one to the other is discontinuous. If the solid angle $d\Omega$ along the line of sight, at angles ($\theta,\phi$), intersects the optically thin zone then the volume of the disk along the line of sight is larger by $\sec \theta$ compared to that along a direction parallel to the disk axis. But if the zone were optically thick then only neutrinos from the material within one scattering length of the disk `surface' can be `seen', a distance that does not vary with the observation angle. Hence the differential number surface fluxes in equation (\ref{eq:ann_rate}), which are defined by `collapsing' the disk's volume flux along a direction parallel to the disk axis, need to take into account this effect. We adopt $S_{\nu}(\theta_{\nu},\phi_{\nu}) = \Theta(a_{\nu}-r_{\nu})+\Theta(r_{\nu}-a_{\nu})\sec \theta_{\nu}$, where $\Theta$ is the Heaviside step function, and similarly for the antineutrinos. 

The limits on the angle integrals are functions of both the position $(r,z)$ and the inner and outer radii of the disk, $R_{in}$ and $R_{out}$. We shall only consider the case for which $r<R_{out}$ so that the limits on $\phi$ are $0$ and $2\pi$. The limits for $\theta$ are then: 
\begin{eqnarray} 
z\,\tan \theta_{min} & = & 
 \left\{\begin{array}{ll}
    \sqrt{ R_{in}^{2}-r^2\,\sin^{2}\phi}  - r\cos\phi,  
                                & \rmn{when}\ r <R_{in}, \\
          0,                    & \rmn{when}\ R_{in}<r<R_{out}, 
  \end{array}\right. \\
z\,\tan \theta_{max} & = & \sqrt{R_{out}^{2}-r^2\,\sin^{2}\phi} - r\cos\phi . 
\end{eqnarray}


\subsection{Neutrino Absorption}

Absorption of neutrinos and antineutrinos by the nucleons, reactions (\ref{reaction:nunabs}) and (\ref{reaction:nubarpabs}), will transfer both momentum and energy to the electrons, positrons and nucleons. If we select a position $(r,z)$ at which to evaluate the transfer from neutrinos emerging from $(\phi,\theta)$ then the momentum transfer occurs along the unit vector ${\bf k}'$. These vectors and angles are illustrated in figure (\ref{fig:geometry}). In terms of ${\bf i,j,k}$ the unit vectors ${\bf i}',{\bf j}',{\bf k}'$ are 
\begin{eqnarray}
{\bf i}' & = & - \cos\phi\cos\theta\, {\bf i} - \sin\phi\cos\theta\, {\bf j} - \sin\theta\, {\bf k}, \\
{\bf j}' & = & \sin\phi\, {\bf i} - \cos\phi\,{\bf j}, \\
{\bf k}' & = & - \cos\phi\sin\theta\, {\bf i} - \sin\phi\sin\theta\, {\bf j} + \cos\,\theta\, {\bf k}
\end{eqnarray}
where ${\bf i}$ is a radial unit vector in the plane of the disk and ${\bf k}$ the axial unit vector. 
Due to the cylindrical symmetry the integration over $\phi$ can only leave the ${\bf i}$ and ${\bf k}$ components of ${\bf k}'$ as non-zero. Thus we obtain an expressions for the rate of total momentum and energy transfer to all particles, ${\bf F}^{(abs)},W^{(abs)}$ at any given position $(r,z)$ per proton or neutron of the form 
\begin{eqnarray}
{\bf F}^{(abs)} & = & \int d\Omega_{\nu} \left[ -\sin\theta\,\cos\phi\, {\bf i} + \cos\theta\, {\bf k} \right] 
\int dE_{\nu} \; \frac{f_{\nu}(r_{\nu},E_{\nu})}{4\pi} \;S_{\nu}(\theta_{\nu},\phi_{\nu})\,E_{\nu}\, \sigma_{\nu N}^{(abs)} , \\
W^{(abs)} & = & \int d\Omega_{\nu} \int dE_{\nu}\, \frac{f_{\nu}(r_{\nu},E_{\nu})}{4\pi} \;S_{\nu}(\theta_{\nu},\phi_{\nu})\, E_{\nu} \, \sigma_{\nu N}^{(abs)}. 
\end{eqnarray}
The cross sections are given by
\begin{eqnarray}
\sigma_{\nu n}^{(abs)} & = & \sigma_{0}\;\left(\frac{1+3\,g_{A}^{2}}{4}\right)\;\left(\frac{E_{\nu}+\Delta}{m_{e}^{2}}\right)^{2} \;\sqrt{1 -\left(\frac{m_{e}}{E_{\nu}+\Delta}\right)^{2} }\;W_{M} \\
\sigma_{\bar{\nu} p}^{(abs)} & = & \sigma_{0}\;\left(\frac{1+3\,g_{A}^{2}}{4}\right)\;\left(\frac{E_{\bar{\nu}}-\Delta}{m_{e}^{2}}\right)^{2} \;\sqrt{1 -\left(\frac{m_{e}}{E_{\bar{\nu}}-\Delta}\right)^{2} }\;W_{\bar{M}} 
\end{eqnarray}
with $g_{A}$ the axial-vector coupling constant and $\Delta$ is the neutron-proton mass difference. In these cross sections $W_{M},W_{\bar{M}}$ are two weak magnetism corrections given by \citet{H2002}
\begin{eqnarray}
W_{M} & = & 1 + 1.1 \frac{E_{\nu}}{m_{n}} \\
W_{\bar{M}} & = & 1 - 7.1 \frac{E_{\bar{\nu}}}{m_{n}}. 
\end{eqnarray}
The limits on the angular integrals are the same as in $\nu + \bar{\nu}$ annihilation. 


\subsection{Neutrino Scattering}

The scattering of neutrinos and antineutrinos by electrons, protons and neutrons, reactions (\ref{reaction:nue}) through (\ref{reaction:nun}), will also transfer momentum and energy. The added complication of course is that, in addition to the incoming neutrino energy and angle integrals, one must also integrate over the outgoing angles. However due to the cylindrical symmetry any ${\bf i}'$ and ${\bf j}'$ terms must vanish leaving only the ${\bf k}'$ component. Again, due to the cylindrical symmetry, there can be no momentum flow in the ${\bf j}$ direction. This leads to momentum and energy transfer expressions due to scattering of 
\begin{eqnarray}
{\bf F} & = & \int d\Omega_{\nu} \left[ -\sin\theta\,\cos\phi\, {\bf i} + \cos\theta\, {\bf k} \right] \int dE_{\nu} \frac{f_{\nu}(r_{\nu},E_{\nu})}{4\pi}\;S_{\nu}(\theta_{\nu},\phi_{\nu})\, \int d\Omega'_{\nu}\; p_{{\bf k}'}\; \frac{d\sigma}{d\Omega'_{\nu}}, \label{eq:Fscat} \\
W & = & \int d\Omega_{\nu} \int dE_{\nu} \frac{f_{\nu}(r_{\nu},E_{\nu})}{4\pi} \;S_{\nu}(\theta_{\nu},\phi_{\nu})\, \int d\Omega'_{\nu}\;T\;\frac{d\sigma}{d\Omega'_{\nu}}, \label{eq:Wscat}
\end{eqnarray}
with 
\begin{eqnarray}
p_{{\bf k}'} & = & \frac{E_{\nu}\,(E_{\nu}+M)\,(1-\cos\theta_{\nu}')}{M+E_{\nu}(1-\cos\theta_{\nu}')}, \\
T & = & \frac{E^{2}_{\nu}\,(1-\cos\theta_{\nu}')}{M+E_{\nu}(1-\cos\theta_{\nu}')} \label{eq:T}
\end{eqnarray}
where $M$ is the mass of the scattered particle and $\theta_{\nu}'$ is the neutrino's scattering angle relative to ${\bf k}'$. These expressions show that the momentum transfer is always $\sim E_{\nu}$, whatever the mass of the scattered particle, but that the energy transferred becomes very small if $M \gg E_{\nu}$. 


\subsubsection{Neutrino-Electron Scattering}

The cross section for electron-neutrino scattering is again well known \citep[see][for example]{B1987,HH1989,P2003} and usually expressed in the frame of reference where the electron is initially at rest as
\begin{equation}
\frac{d\sigma_{\nu e}}{dT_{e}} = \frac{\sigma_{0}}{8\,m_{e}} \left( (C_{V}+C_{A})^{2} + (C_{V}-C_{A})^{2}\,\left(1-\frac{T_{e}}{E_{\nu}}\right)^{2} -(C_{V}^{2}-C_{A}^{2})\,\frac{m_{e}\,T_{e}}{E_{\nu}^{2}} \right) \label{eq:sigescatt}
\end{equation}
where $T_{e}$ is the kinetic energy of the recoiling electron. The transformation to $d\sigma_{\nu e}/d\Omega_{\nu}$, the quantity needed for equations (\ref{eq:Fscat}) and (\ref{eq:Wscat}), is achieved by multiplication of (\ref{eq:sigescatt}) by $(2\pi)^{-1}\,dT_{e}/d(\cos \theta'_{\nu})$ with $T_{e}(\cos \theta'_{\nu})$ given by equation (\ref{eq:T}). For electron-antineutrino scattering $C_{A} \rightarrow -C_{A}$. We use the approximation that the electrons are initially at rest, however, if the temperature of the material becomes of order an MeV, or it acquires large velocities, then corrections due to the finite electron momentum would need to be taken into account.


\subsubsection{Neutrino-Nucleon Scattering} 

The differential cross sections for these processes are given by \citet{B2001}
\begin{eqnarray}
\frac{d\sigma_{\nu n}}{d\Omega_{\nu}} & = &  \frac{\sigma_{0}}{16\pi} \left(\frac{E_{\nu}}{m_{e}}\right)^{2} \; \left(\frac{1+3\,g_{A}^{2}}{4}\right) \left( 1+ \delta_{n}\,\cos \theta'_{\nu} \right), \\
\frac{d\sigma_{\nu p}}{d\Omega_{\nu}} & = &  \frac{\sigma_{0}}{16\pi} \left(\frac{E_{\nu}}{m_{e}}\right)^{2} \; \left[(C_{V}-1)^{2} + 3\,g_{A}^{2}(C_{A}-1)^{2}\right] \left( 1+ \delta_{p}\,\cos \theta'_{\nu} \right) 
\end{eqnarray}
where 
\begin{eqnarray}
\delta_{n} & = & \frac{1-g_{A}^{2}}{1+3 g_{A}^{2}}, \\
\delta_{p} & = & \frac{(C_{V}-1)^{2} - g_{A}^{2}(C_{A}-1)^{2}}{(C_{V}-1)^{2} + 3\,g_{A}^{2}(C_{A}-1)^{2}}.
\end{eqnarray}
The cross sections for antineutrino scattering are identical except for the inconsequential transform of $g_{A} \rightarrow -g_{A}$. 


\section{Results} \label{sec:results}

Table (\ref{table:luminosities}) shows the neutrino plus antineutrino luminosity coming from each of the disk models. Our luminosities are slightly smaller than those \citet{pwf99} for the same models by roughly a factor of 1.6. From examination of the table the basic trends with variations of the disk parameters show that the luminosity increases with both the accretion rate and spin parameter but that larger black hole masses lead to smaller luminosities. For accretion rates as large as $1 \,\rmn{M_{\sun}\, s^{-1}}$, the total luminosity from the accretion disk is as high as $10^{53}\rmn{erg\,s^{-1}}$. However, for low accretion rate disks, such as $0.01 \,\rmn{M_{\sun} \,s^{-1}}$, the luminosity in neutrinos plus antineutrinos, $L_{\nu}$, coming from the disk can be as low as $10^{49}\rmn{erg\,s^{-1}}$. The fraction of the rest mass energy radiated from the disk, $L_{\nu}/\dot{m}$, also shown in table (\ref{table:luminosities}), can be as large as $\sim 10$ per cent when the spin parameter and/or accretion rate are large. There is also clearly some interdependance amongst the disk parameters with much larger increases seen in $L_{\nu}$ of the $\alpha = 0.1$ thin-disk models when $\dot{m}$ is increased compared to those models with $\alpha = 0.01$. Similarly, the reduction in luminosity due to increased viscosity is apparently more pronounced when the accretion rate is low compared to those cases with moderate accretion rates. The details of the model are also seen to have an effect (models \modtenpwf~ and \modonedpn) with a doubling of the neutrino luminosity for those that include trapping as compared to the thin disk model.  

Table \ref{table:luminosities} also shows the total neutrino-antineutrino annihilation luminosity, $L_{\nu\bar{\nu}}$, integrated over the volume above and below the disk. The total power deposited when the accretion rate and/or the spin parameter is large can be considerable, approaching $\sim 5 \times 10^{51}\;\rmn{erg\,s^{-1}}$ for $\dot{m}=10\,\rmn{M_{\sun}\,s^{-1}}$. In comparison with \citet{pwf99} our results for the same models are smaller by roughly a factor of 5-6. The difference is mainly due to the fact that they compute the neutrino-antineutrino annihilation rate using Eq.~(6-1) in \citet{pwf99} while we use Eq.~(\ref{eq:ann_rate}). By comparing the models we notice that the annihilation luminosity varies more rapidly with changes in the disk parameters than $L_{\nu}$. Increasing the spin parameter from 0 to 0.5 produces a factor of $\sim2$ increase in the neutrino luminosity but a factor of $\sim8$ in the annihilation luminosity. Likewise the increase in $L_{\nu\bar{\nu}}$ in the thin disk models as the accretion rate grows is much more pronounced, in some models from factors of $\sim 200$ for $L_{\nu}$ to factors of $~10^{5}$ for $L_{\nu\bar{\nu}}$. The parameter interdependance noted above for $L_{\nu}$ is again seen in $L_{\nu\bar{\nu}}$. Interestingly the inclusion of trapping into the model (models \modtenpwf~ and \modonedpn) seems to reduce the annihilation luminosity slightly. The efficiency with which the neutrino energy is tapped, $L_{\nu\bar{\nu}}/L_{\nu}$, is also shown in the table. Again this displays a greater variation than does $L_{\nu}$ reaching $\sim 0.3$ per cent for large accretion rates and spin parameters but can be as small as $\sim 4 \times 10^{-7}$ if the disk is viscous, the accretion rate is small, or the black hole mass is large.

\begin{figure}
\includegraphics[width=7in]{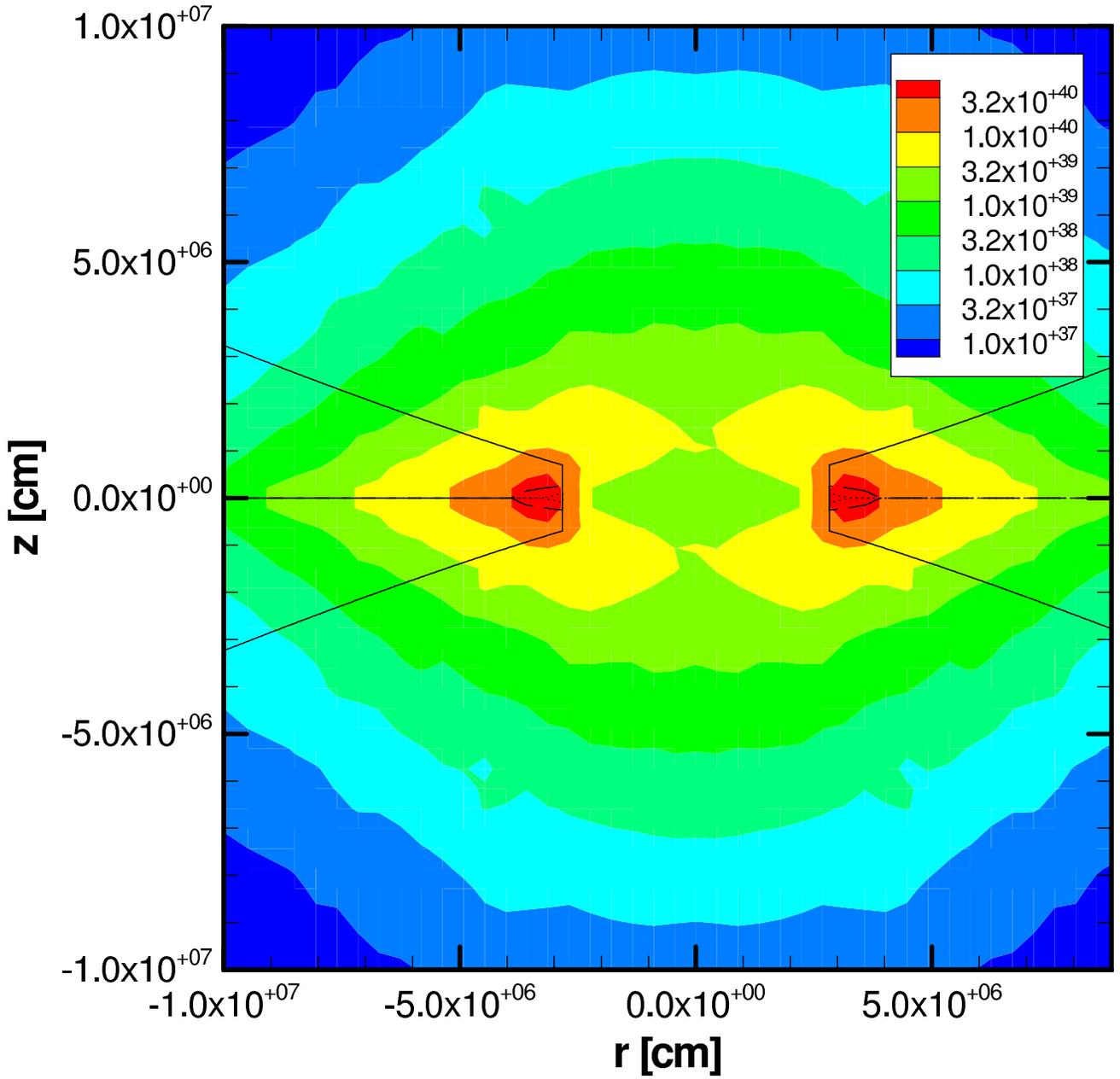}
\caption{Shows the energy deposited due to neutrino-antineutrino annihilation in units of $\rmn{ eV\,cm^{-3}\,s^{-1}}$. The x-axis is distance radially along the disk, while the y axis is vertical distance above the disk. This model is DPN $\dot{m} = 1\,\rmn{M_{\sun}\,s^{-1}}$. The density scale height of the disk is shown as a solid line while the trapped regions of the neutrinos and antineutrinos are shown as dashed and dotted lines respectively.}
\label{fig:nuann}
\end{figure}

Focusing upon model \modonedpn, we show in figure (\ref{fig:nuann}) the $\nu-\bar{\nu}$ annihilation power as a function of $r$ and $z$. The result is very similar to that of \citet{ruf97} for neutron star mergers. The bulk of neutrino annihilation occurs in an ellipsoidal region extending upwards to $z \sim 40\;\rmn{km}$ and outwards to $r \sim 90\;\rmn{km}$ peaked in the vicinity of the strongest neutrino-antineutrino flux namely the innermost regions of the disk.  The oblateness is the product of both the diminishing flux as one moves away from the brightest zones on the disk \emph{and} the cross section's dependence upon the angle between the neutrino trajectories. Along the z axis $dL_{\nu\bar{\nu}}/dV$ varies as $dL_{\nu\bar{\nu}}/dV \propto z / ( z_{\star}^{2}+z^{2})^{7/2}$ (c.f. \citet{ruf97} with $z_{\star}$ a constant equal to $\sim 40\;\rmn{km}$ for this particular model\footnote{A least square fit of our results to the equation $dL_{\nu\bar{\nu}}/dV = (C_{0} + C_{1}\,z^{m} ) / ( z_{\star}^{2}+z^{2})^{n}$ with $m,\,n, z_{\star}, C_{0}$ and $C_{0}$ all adjustable and, following the numerical methods used in the computation of equation (\ref{fig:nuann}), assignation of an `error' proportional to the computed values, gives $\{m,\,n, z_{\star}\}=\{0.87,3.77,42.6\;\rmn{km} \}$ with the peak at $14.9\;\rmn{km}$. \label{foot:Wfit}}. Along the disk axis the peak power deposition occurs at $z_{\star}/\sqrt{6} \sim 16\;\rmn{km}$ but note that, according to the figure, the power deposited at this point is more than an order of magnitude smaller than that immediately above the innermost portion of the disk. This fact has also been noted by \citet{geo03} who used a thin disk model taking into account general relativistic effects. The electron positron pairs will thermalize locally so one should expect a large temperature and pressure difference between the `hot spots' immediately around the innermost portion of the disk and the z axis. 

In contrast, the annihilation power along the z axis for the thin disk models, \modonepwf-\modtenpwf, varies as $dL_{\nu\bar{\nu}}/dV \propto 1 / ( z_{\star}^{2}+z^{2})^{3}$ and in model \modtendpn~ this changes to $dL_{\nu\bar{\nu}}/dV \propto z^{2} / ( z_{\star}^{2}+z^{2})^{4}$. The difference between models \modonedpn~ and \modtendpn~ lie in the contribution of the thick and thin portion of the disk to the total antineutrino flux at any given point above the disk: $a_{\bar{\nu}}= 33.6\;\rmn{km}$ in model \modonedpn~ vis-a-vis $a_{\bar{\nu}}= 158.1\;\rmn{km}$ in model \modtendpn. At large distances, $z \gg z_{\star}$, all the models seem to follow a similar scaling with z, it is the structure close to the black hole that is different. Neutrino opacity not only affects the structure of the disk and the total annihilation power but it also manifest itself in \emph{where} this power is deposited and, consequently, the expected temperature and pressure gradients produced upon local electron-positron thermalization will be altered. 

\begin{figure}
\includegraphics[width=7in]{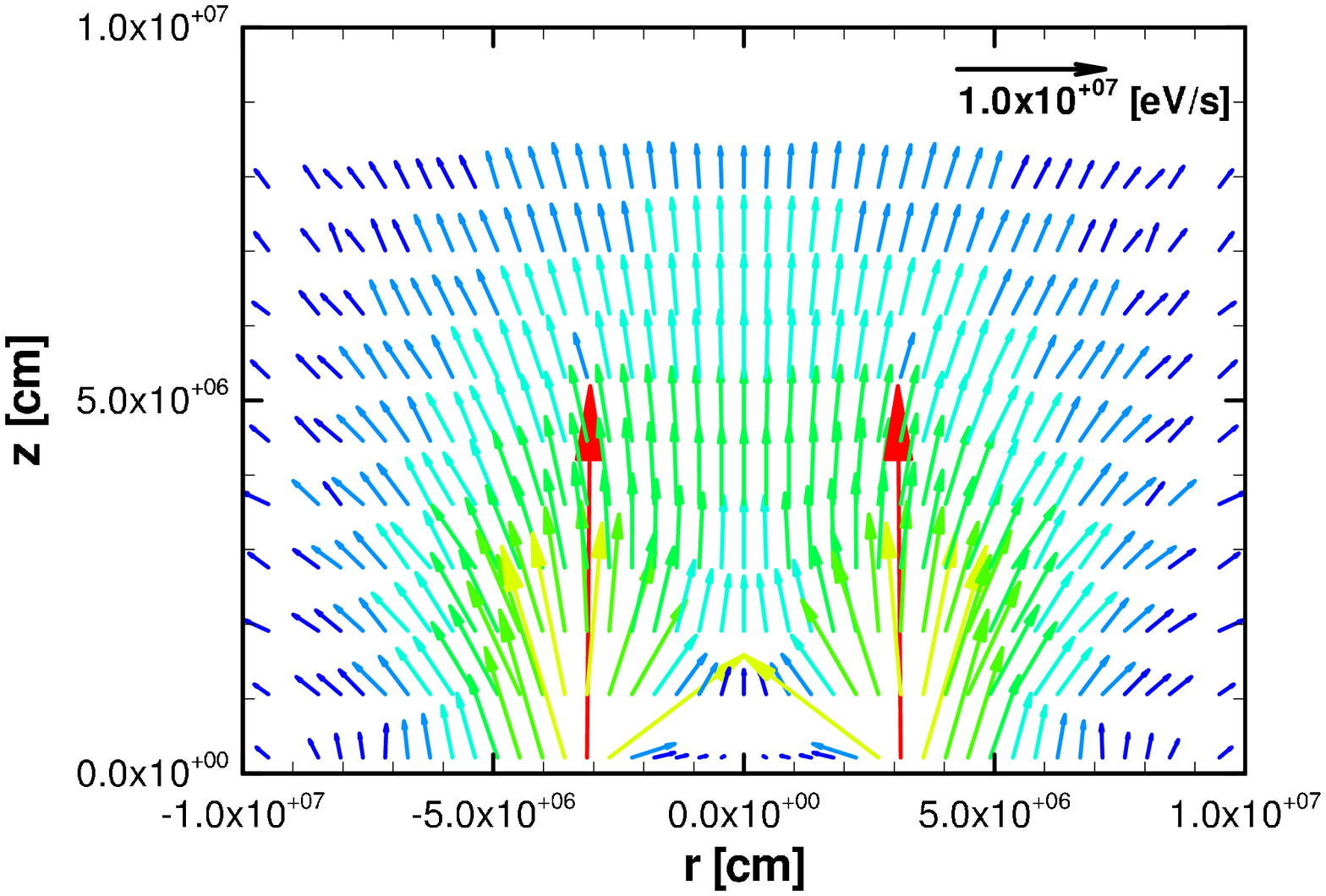}
\caption{Shows the momentum transfer from neutrino-electron scattering at all points above the accretion disk in the DPN $\dot{m} = 1\,\rmn{M_{\sun}\,s^{-1}}$ model. The x and y axis are the same as in Fig. \ref{fig:nuann}.}
\label{fig:escatt}
\end{figure}

Table \ref{table:peaks} shows the maximum momentum transfer along the z axis for each of the eleven models due to proton and neutron absorption and then neutron, proton and electron scattering. We give the momentum along the z-axis because this is the direction in which the jet would emerge but we should note that the momentum transfer at other points above the disk is larger than along the axis, see Fig. (\ref{fig:escatt}). 
The peak momentum transfer varies as strongly with the disk parameters as does the neutrino luminosity and neutrino-antineutrino annihilation power in table \ref{table:luminosities}. A comparison between the entries within each model shows that the momentum transferred due to neutrino absorption is larger than that in the scattering reactions but if the material along the z axis is primarily leptonic then the scattering of neutrinos by electron/positron will dominate. Across different models we see increases in $F_{k}$ as $\dot{m}$ grows except for very large $\dot{m}$, models \modonedpn~ and \modtendpn, where the trend is reversed for the neutrinos; larger momentum transfer with increases in the spin parameter and inverse correlation with the black hole mass. The effects of the viscosity are mixed with smaller transfers when the accretion rate is small but larger differences at moderate rates. Whether trapping is included also seems to play a role with the momentum transfer along the disk axis being larger by a factor of $\sim 3-4$.

In Fig. (\ref{fig:escatt}) we show the momentum transfer from electron-neutrino scattering as a function of $r$ and $z$ again for our model \modonedpn. At large distances from the black hole the momentum transfer is purely radial but as we approach the black hole the non-spherical nature of the source becomes apparent. The largest momentum transfer occurs in the vicinity of the inner region of the disk where the neutrino fluxes are largest and there is an oblate, spheroidal region, extending up to $z \sim 36\;\rmn{km}$ for this model, in which the momentum transfer is directed towards the disk axis. The momentum transferred in the z direction also reaches its maximum value (listed in table \ref{table:peaks}) at this point which is, therefore, further out than the peak in $dL_{\nu\bar{\nu}}/dV$. 

A rough fit to the momentum transferred due to electron-neutrino scattering along the disk axis\footnote{A least square fit, as in discussed in footnote \ref{foot:Wfit}, gives $F_{k} \propto z^{1.87}/(z_{\star}^{2} + z^{2})^{1.9}$ with $z_{\star} =37.4\;\rmn{km}$.} in model \modonedpn~ is $F_{k} \propto z^{2}/(z_{\star}^{2} + z^{2})^{2}$, whereas antineutrino-electron scattering is closer to\footnote{A least square fit gives $F_{k} \propto z^{1.21}/(z_{\star}^{2} + z^{2})^{1.58}$ with $z_{\star} =40.6\;\rmn{km}$.} $F_{k} \propto z/(z_{\star}^{2} + z^{2})^{3/2}$. Again the difference is due to the rather small extent of the trapped antineutrino zone. Both neutrino-electron and antineutrino-electron scattering in all the thin disk models follow the $z/(z_{\star}^{2} + z^{2})^{3/2}$ pattern but in model \modtendpn~ it is $z^{2}/(z_{\star}^{2} + z^{2})^{2}$ that is the better match. Thus trapping, and its extent, also appear to alter the details of where, how much, and in which direction electrons gain momentum especially in the vicinity of the black hole.  

\begin{figure}
\includegraphics[width=7in]{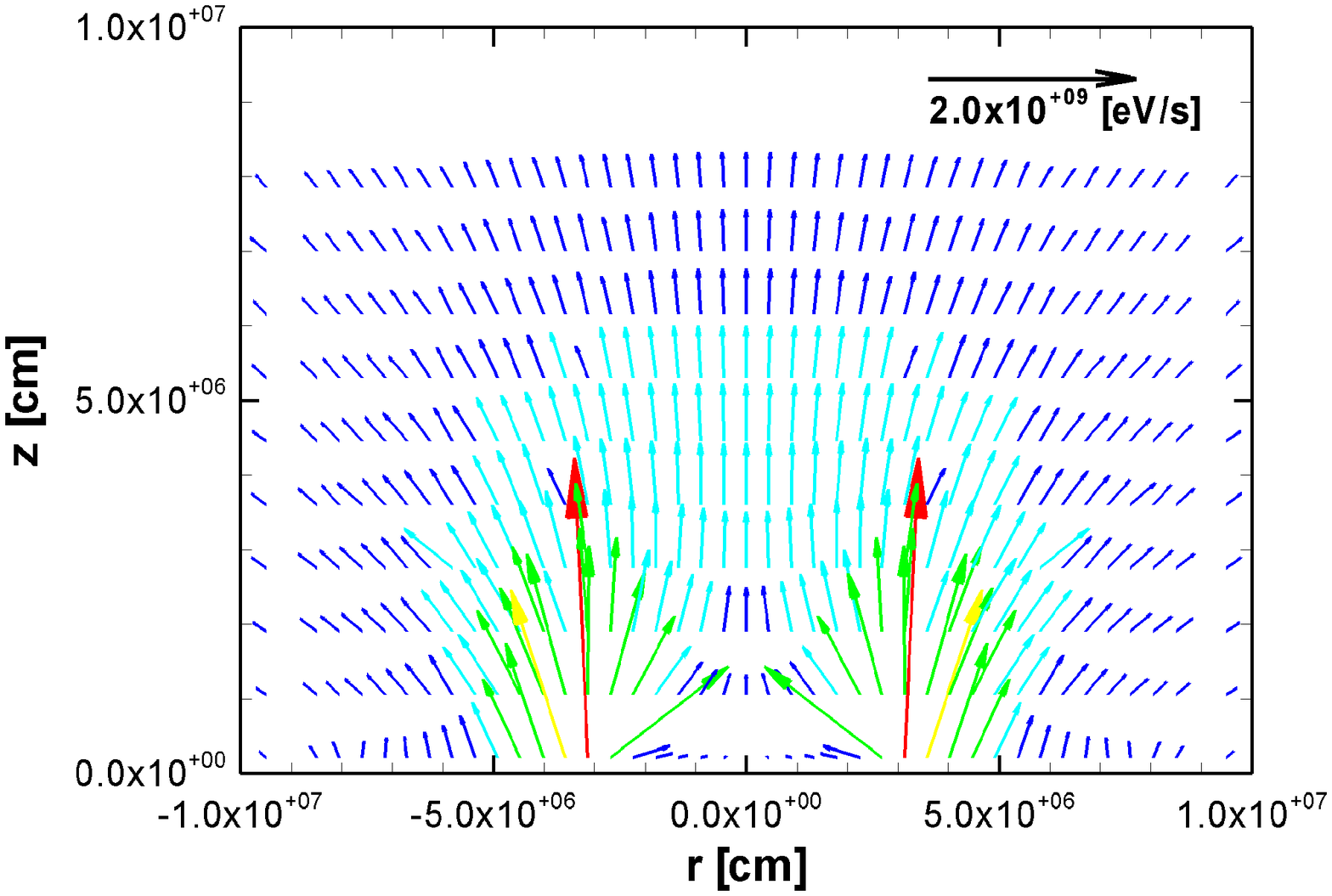}
\caption{Shows the momentum transfer from neutrino-proton scattering at all points above the accretion disk in the DPN $\dot{m} = 1 \,\rmn{M_{\sun}\,s^{-1}}$ model. The x and y axis are the same as in Fig. \ref{fig:nuann}.}
\label{fig:pscatt}
\end{figure}

Finally in Fig. (\ref{fig:pscatt}) we show the momentum transfer from proton-neutrino scattering as a function of $r$ and $z$ again for our model \modonedpn. The same basic structure is seen and, again, the momentum transferred in the z direction reaches its maximum value at $z \sim 36\;\rmn{km}$. However note that the momentum transferred is somewhat larger than that in $e-\nu$ scattering seen in figure (\ref{fig:escatt}). The reason lies in the cross sections for each: for neutrino-electron scattering $\sigma$ varies linearly with $E_{\nu}/m_{e}$ but in proton-neutrino scattering we have $\sigma \propto E^{2}_{\nu}/m^{2}_{e}$. For neutrino energies of, say, $\sim 5\;\rmn{MeV}$ the number of  proton-neutrino scatterings is larger by, roughly, an order of magnitude. As we mentioned above, the momentum transfer per scattering for each process is comparable so the net momentum transfer, the scattering rate multiplied by the average momentum transferred, is greater for proton-neutrino than for electron-neutrino scattering. This would change if the electrons had kinetic energy or temperature of order $\sim \;\rmn{MeV}$ also. 


\subsection{Critical Densities}

The energy deposited into electron/positron pairs and the scattering of neutrinos will retard the inward motion of material towards the black hole. Following the arguments in \citet{fry03} for the case of material falling along the disk axis, as a mass element, with density $\rho$, falls towards the black hole it experiences an acceleration due to  gravity, $F_{G}$, and a deceleration from both scattering plus absorption, $F_{\nu}$, and from neutrino annihilation, $F_{\nu\bar{\nu}}$. If the mass elements freefalls from infinity then after it has fallen to the height $z$ the change in kinetic energy is 
\begin{equation}
\frac{\rho}{2}\,v^{2}(z) = \int_{z}^{\infty}\,dz\, ( F_{G} + F_{\nu} + F_{\nu\bar{\nu}} ) \label{eq:deltaT}
\end{equation}
The force of gravity is simply $F_{G} = - G_{N}\,M_{BH}\,\rho /z^{2}$, we take the force due to neutrino-antineutrino annihilation as $F_{\nu\bar{\nu}} = 1/c\; dL_{\nu\bar{\nu}}/dV$, and the force due to scattering and absorption is 
\begin{equation}
F_{\nu} = n_{n}\,\left( F_{n}^{(abs)} + F_{n\nu} + F_{n\bar{\nu}} \right) + n_{p}\,\left( F_{p}^{(abs)} + F_{p\nu} + F_{p\bar{\nu}} \right) + n_{e}\,\left(F_{e\nu} + F_{e\bar{\nu}} \right)
\end{equation}
where $n_{i}$ are the number densities. Using $n_{e} = n_{p}$, $\rho = (n_{n}+n_{p})\,m_{u}$, where $m_{u}$ is the atomic mass unit, and introducing the neutron fraction $Y= n_{n}/(n_{n}+n_{p})$, then 
\begin{equation}
F_{\nu} = \frac{\rho}{m_{u}}\;\big[ Y\,F_{n} + (1-Y)\,(F_{p}+F_{e})\big] 
\end{equation}
where we introduce $F_{n}$, $F_{p}$ and $F_{e}$ to represent the total momentum transfer rates. Both $F_{G}$ and $F_{\nu}$ are proportional to $\rho$ and so at some value, $\rho_{0}$, the right hand side of equation (\ref{eq:deltaT}) vanishes. Mass elements with densities equal to $\rho_{0}$ will then turn around at $z$ and be ejected; if the density is smaller than $\rho_{0}$ then the motion will have been reversed at some height greater than $z$; if it is larger then the mass element will pass by $z$. Solving equation (\ref{eq:deltaT}) for this density gives 
\begin{equation}
\rho_{0} = \int_{z}^{\infty}\,dz\,\frac{1}{c}\;\frac{dL_{\nu\bar{\nu}}}{dV} \;\bigg/\; \int_{z}^{\infty}\,dz\,\left[ \frac{G_{N}\,M_{BH}}{z^{2}} - \frac{\left[F_{n}\,Y + (F_{p}+F_{e})(1-Y)\right]}{m_{u}} \right].
\end{equation}
Clearly $\rho_{0}$ is a function of $z$ and has a maximum value, $\rho_{\star}$, we name the critical density. Material with a density larger than $\rho_{\star}$ cannot be reversed at \emph{any} value of $z$ so must be accreted into the black hole. Using $Y_{n}=0.5$ these critical densities are listed in table (\ref{table:densities}). When a comparison can be made with the results in \citet{fry03} our results are somewhat smaller. 


\section{Discussion and Conclusions}

The accretion disk model of the central engine of GRBs is an attractive one whose ultimate success will rely on the efficient extraction of the energy and momentum of the neutrinos emitted from the disk. The non-spherical geometry of the source makes this easier to achieve than in the context of a spherical source such as in a Type II supernova. The neutrinos are essentially emitted from a ring: a configuration that allows neutrinos to approach one another at larger interaction angles, $\gamma$, with the concomitant increase in center of mass energy. The bulk of neutrino annihilation occurs in an oblate spheroidal region, extending upwards to $z \sim 40 \;\rmn{km}$ and outward to $r \sim 90 \;\rmn{km}$ for our model \modonedpn, with the highest deposited power in the vicinity of the inner edge of the disk. In comparison with other azimuths, the power deposited along the z axis is a minimum. The electron/positron pairs would locally thermalize so one should expect a large pressure gradient directed toward the disk axis. In addition neutrino absorption and scattering creates a region, close to the black hole, in which the momentum transfer is also directed towards the disk axis. Trapping of neutrinos and antineutrinos within the disk leads to significant differences in the energy and momentum deposition structure particularly in the region close to the black hole. This may lead to noticeable changes in the explosion dynamics. 

These results suggest a scenario whereby the electron/positron pairs created due to neutrino annihilation would, at least in part, move initially `inward and upward' facing little resistance from the $e^{+}/e^{-}$ pairs created along the z axis and accelerated by neutrino scattering from behind. At the disk axis the converging $e^{+}/e^{-}$ gas would collide, further increasing the internal energy, and then the hot, lepton rich gas flowing rapidly along the z axis. The movement of the $e^{+}/e^{-}$ gas created by neutrino annihilation would have consequences for an explosion. The results in table \ref{table:densities} used only the power deposited along the disk axis to calculate the density of material that could be ejected but the movement of the $e^{+}/e^{-}$ pairs from the `hot spots' around the inner disk would suggest that these densities may be too small. The collision at the disk axis of the $e^{+}/e^{-}$ gas from around the torus from would raise the internal energy and, furthermore, the increase in the number density would also increase the neutrino opacity leading to a greater transfer of their momentum. Larger critical densities would also 
imply the more prompt formation of a jet. 


\section*{Acknowledgments}

The authors would like to thank Asif Ud-Doula and Chris Fryer for useful discussions. This work is supported at NCSU by the DOE under grant DE-FG02-02ER41216.


\clearpage


\begin{table}
\caption{Disk Models}
\label{table:models}
\begin{tabular}{@{}cccccccc} 
\hline
Model & Disk Calculation$^{a}$ & $\dot{m}$ & $\alpha$ & $a$ & $M_{BH}$ & $a_{\nu}$& $a_{\bar{\nu}}$ \\ 
      & & $[\rmn{M_{\sun}\,s^{-1}}]$& & & $[\rmn{M_{\sun}}]$ & $[\rmn{km}]$ & $[\rmn{km}]$ \\
\hline
\modonepwf & PWF & 0.01 & 0.1 & 0 & 3 & & \\
\modtwopwf & PWF & 0.01 & 0.01 & 0 & 3 & & \\
\modthreepwf & PWF & 0.01 & 0.1 & 0.5 & 3 & & \\
\modfourpwf & PWF & 0.01 & 0.01 & 0 & 10 & & \\
\modfivepwf & PWF & 0.1 & 0.1 & 0 & 3 & & \\
\modsixpwf & PWF & 0.1 & 0.01 & 0 & 3 & & \\
\modsevenpwf & PWF & 0.1 & 0.1 & 0.5 & 3 & & \\
\modninepwf$^{b}$ & PWF & 0.1 & 0.1 & 0.95 & 6 & & \\
\modtenpwf$^{b}$ & PWF & 1.0 & 0.1 & 0 & 3 & & \\
\modonedpn & DPN & 1.0 & 0.1 & 0 & 3 & 72.6 & 33.6 \\
\modtendpn & DPN & 10.0 & 0.1 & 0 & 3 & 240.2 & 158.1 \\ 
\hline
\end{tabular}\\
$^{a}\;$ The acronyms PWF and DPN denote models from \citet{pwf99} and \citet{dim02} respectively.\\
$^{b}\;$ It is likely that (unaccounted for) neutrino trapping is important in these models. 
\end{table}

\begin{table*}
\begin{minipage}{7in}
\caption{Luminosities and Efficiencies}
\label{table:luminosities}
\begin{tabular}{@{}ccccc} 
\hline
Model & $L_{\nu}$ & $L_\nu / \dot{m}$ & $L_{\nu\bar{\nu}}$ & $L_{\nu \bar{\nu}} / L_\nu $ \\ 
      & $[10^{51}\rmn{erg\,s^{-1}}]$ & & $[10^{51} \rmn{erg\,s^{-1}}]$ & \\
\hline
\modonepwf & 0.0094 & $5.28 \times 10^{-4}$&  $4.01 \times 10^{-9}$ & $4.25 \times 10^{-7}$\\
\modtwopwf & 0.396 & $2.22 \times 10^{-2}$ &  $2.90 \times 10^{-6}$  & $7.31 \times 10^{-6}$\\
\modthreepwf & 0.0207 & $1.16 \times 10^{-3}$ &  $3.24 \times 10^{-8}$ & $1.56 \times 10^{-6}$\\
\modfourpwf & 0.0301 & $1.68 \times 10^{-3}$ &  $5.28 \times 10^{-9}$ & $1.75 \times 10^{-7}$\\
\modfivepwf & 2.35 & $1.31 \times 10^{-2}$ &  $4.13 \times 10^{-4}$  & $1.76 \times 10^{-4}$\\
\modsixpwf & 2.14 & $1.20 \times 10^{-2}$ &  $4.05 \times 10^{-4}$  & $1.90 \times 10^{-4}$\\
\modsevenpwf & 5.34 & $2.99 \times 10^{-2}$ &  $3.58 \times 10^{-3}$  & $6.70 \times 10^{-4}$\\
\modninepwf & 15.3 & $8.57 \times 10^{-2}$ & $4.27 \times 10^{-2}$   & $2.79 \times 10^{-3}$\\
\modtenpwf & 54.3 & $3.03 \times 10^{-2}$ & $0.124$                & $2.29 \times 10^{-3}$\\
\modonedpn & 109 & $6.13 \times 10^{-2}$ &  $0.117$                & $1.07 \times 10^{-3}$\\
\modtendpn & 1674 & $9.37 \times 10^{-2}$ &  $4.66$                & $2.79 \times 10^{-3}$\\
\hline
\end{tabular}
\end{minipage}
\end{table*}

\begin{table*}
\begin{minipage}{7in}
\caption{Peak Momentum transfer per target particle along z axis}
\label{table:peaks}
\begin{tabular}{@{}ccccccccc} 
\hline
Model & ${\bf F}^{(abs)}_{n}$ & ${\bf F}^{(abs)}_{p}$ & ${\bf F}_{n\nu}$ & ${\bf F}_{n\bar{\nu}}$ & ${\bf F}_{p\nu}$ & ${\bf F}_{p\bar{\nu}}$ & ${\bf F}_{e\nu}$ & ${\bf F}_{e\bar{\nu}}$ \\ 
& $[\rmn{g\,cm\,s^{-2}}]$ & $[\rmn{g\,cm\,s^{-2}}]$  & $[\rmn{g\,cm\,s^{-2}}]$  & $[\rmn{g\,cm\,s^{-2}}]$  & $[\rmn{ g\,cm\,s^{-2}}]$  & $[\rmn{g\,cm\,s^{-2}}]$  & $[\rmn{g\,cm\,s^{-2}}]$  & $[\rmn{g\,cm\,s^{-2}}]$ \\
\hline
\modonepwf & $1.68 \times 10^{-17}$ & $1.48 \times 10^{-17}$ & $3.71 \times 10^{-18}$ & $5.27 \times 10^{-18}$ 
& $3.30 \times 10^{-18}$ & $4.69 \times 10^{-18}$ & $4.22 \times 10^{-19}$ & $2.38 \times 10^{-18}$ \\
\modtwopwf & $6.15 \times 10^{-17}$ & $3.07 \times 10^{-17}$ & $1.31 \times 10^{-17}$ & $1.08 \times 10^{-17}$ 
& $1.17 \times 10^{-17}$ & $9.67 \times 10^{-18}$ & $1.54 \times 10^{-18}$ & $3.69 \times 10^{-18}$ \\
\modthreepwf & $9.35 \times 10^{-17}$ & $7.54 \times 10^{-17}$ & $2.09 \times 10^{-17}$ & $2.75 \times 10^{-17}$ 
& $1.86 \times 10^{-17}$ & $2.45 \times 10^{-17}$ & $2.36 \times 10^{-18}$ & $1.37 \times 10^{-17}$ \\
\modfourpwf & $2.96 \times 10^{-19}$ & $2.77 \times 10^{-19}$ & $5.93 \times 10^{-20}$ & $1.01 \times 10^{-19}$ 
& $5.29 \times 10^{-20}$ & $9.03 \times 10^{-20}$ & $7.43 \times 10^{-19}$ & $2.63 \times 10^{-20}$ \\
\modfivepwf & $1.17 \times 10^{-14}$ & $7.11 \times 10^{-15}$ & $2.65 \times 10^{-15}$ & $2.80 \times 10^{-15}$ 
& $2.35 \times 10^{-15}$ & $2.49 \times 10^{-15}$ & $3.02 \times 10^{-16}$ & $1.73 \times 10^{-15}$ \\
\modsixpwf & $3.39 \times 10^{-16}$ & $1.27 \times 10^{-16}$ & $7.21 \times 10^{-17}$ & $4.51 \times 10^{-17}$ 
& $6.41 \times 10^{-17}$ & $4.01 \times 10^{-17}$ & $8.50 \times 10^{-18}$ & $1.57 \times 10^{-17}$ \\
\modsevenpwf & $6.20 \times 10^{-14}$ & $3.40 \times 10^{-14}$ & $1.40 \times 10^{-14}$ & $1.42 \times 10^{-14}$ 
& $1.25 \times 10^{-14}$ & $1.26 \times 10^{-14}$ & $1.62 \times 10^{-15}$ & $9.62 \times 10^{-15}$ \\
\modninepwf & $2.97 \times 10^{-13}$ & $1.48 \times 10^{-13}$ & $6.73 \times 10^{-14}$ & $6.79 \times 10^{-14}$ 
& $5.99 \times 10^{-14}$ & $6.04 \times 10^{-14}$ & $7.96 \times 10^{-15}$ & $5.16 \times 10^{-14}$ \\
\modtenpwf & $1.03 \times 10^{-13}$ & $3.03 \times 10^{-14}$ & $2.33 \times 10^{-14}$ & $1.15 \times 10^{-14}$ 
& $2.07 \times 10^{-14}$ & $1.02 \times 10^{-14}$ & $2.63 \times 10^{-15}$ & $6.41 \times 10^{-15}$ \\
\modonedpn & $4.14 \times 10^{-13}$ & $1.04 \times 10^{-13}$ & $9.30 \times 10^{-14}$ & $3.91 \times 10^{-14}$ 
& $8.23 \times 10^{-14}$ & $3.47 \times 10^{-14}$ & $1.05 \times 10^{-14}$ & $2.13 \times 10^{-14}$ \\
\modtendpn & $1.72 \times 10^{-13}$ & $2.62 \times 10^{-12}$ & $3.85 \times 10^{-14}$ & $1.03 \times 10^{-12}$ 
& $3.43 \times 10^{-14}$ & $9.19 \times 10^{-13}$ & $4.38 \times 10^{-15}$ & $6.41 \times 10^{-13}$ \\
\hline
\end{tabular}
\end{minipage}
\end{table*}

\begin{table}
\caption{Critical Densities: the maximum density of material along the $z$ axis whose motion can be reversed.}
\label{table:densities}
\begin{tabular}{@{}cc} 
\hline
Model & $\rho_{\star}$ \\ 
      & $[\rmn{g\,cm^{-3}}]$ \\
\hline
\modonepwf & $9.57 \times 10^{-4}$ \\
\modtwopwf & $0.215$ \\
\modthreepwf & $1.04 \times 10^{-2}$ \\
\modfourpwf & $4.65 \times 10^{-5}$ \\
\modfivepwf & $90.8$ \\
\modsixpwf & $4.06$ \\
\modsevenpwf & $1.05 \times 10^{3}$ \\
\modninepwf & $7.26 \times 10^{3}$ \\
\modtenpwf & $1.51 \times 10^{4}$ \\
\modonedpn & $1.79 \times 10^{4}$ \\
\modtendpn & $8.38 \times 10^{5}$ \\
\hline
\end{tabular}
\end{table}

\clearpage

\bsp

\label{lastpage}

\end{document}